# Observations of a glaciating hole-punch cloud

**Chris Westbrook** *Department of Meteorology, University of Reading*
**Owain Davies** *STFC Chilbolton Facility for Atmospheric and Radio Research*

In his fascinating review of fallstreak holes or 'hole-punch' clouds, David Pedgley (2008) remarked that there were no measurements into the microphysics and dynamics of hole-punch clouds from ground-based remote sensors because the probability of a hole drifting into the instrument sample volume is so small. Here we present some remarkable measurements of exactly that, which we were lucky enough to make at the Chilbolton Observatory in Hampshire.

**What are they?**
These circular (or sometimes cylindrical) holes are formed when aircraft penetrate supercooled liquid cloud layers such as altocumulus. The passage of the aircraft is believed to produce large numbers of ice crystals, which then grow quickly by the Bergeron-Findeison mechanism in the vapour-rich conditions (water-saturated air is supersaturated relative to ice). This growth depletes the cloud of water vapour, causing the liquid droplets to evaporate, and the formation of a hole in the liquid cloud layer with a visible trail of ice crystals falling out beneath. Photographs of these spectacular phenomena are a regular feature in *Weather* (see for example the front cover of March 2009's issue); more detail of the physical mechanisms behind their formation are explored by Pedgley (2008) in his review.

**Observations**
On the 17 April 2007 the Chilbolton area was overcast by stratocumulus for much of the early morning, as indicated by the 0600UTC radiosonde ascent from Larkhill, 25km west of Chilbolton (Figure 1), which shows liquid water saturated layers a few hundred metres thick at approximately 1000m and 2000m altitude. As the morning went on, this low-level cloud broke up somewhat, allowing the ground-based lidars at Chilbolton to see through to a higher level altocumulus layer above. The time series from the 905nm lidar ceilometer (Figure 2, top right panel) shows this altocumulus layer as a thin red strip of strong backscatter ($10^{-4} m^{-1} sr^{-1}$ or greater) between 5400 and 5600m. Lidar measurements are extremely sensitive to liquid clouds, because the concentration of droplets in such clouds is so high, making them very reflective to infrared (and visible) light; they are also optically thick for the same reason, and attenuate the lidar beam over depths of a few hundred metres. The radiosonde ascent confirms the presence of a liquid water saturated layer at this altitude, at a temperature of approximately -20°C, and the droplets in this cloud are therefore supercooled. The air above the cloud layer is drier and more stable, inhibiting further vertical development. Aircraft flying at all levels are often observed in the Chilbolton area, particularly to the West and North. These may come from nearby Popham and Thruxton airfields, but also frequently from further afield. Examination of the time series from an automated camera at Chilbolton does suggest aircraft activity to the West on this particular morning, however the resolution of the digital camera is too coarse to provide any indication of what type.

Although ice crystals were being produced naturally by the altocumulus layer as a small fraction of the droplets freeze, the ceilometer is quite insensitive to these crystals, particularly during the day when the background noise level is much higher because of skylight entering the telescope. These crystals are however highly visible to the second lidar at Chilbolton, a 1.5µm Doppler lidar, and the time series from this instrument is also shown in Figure 2. This instrument points directly at vertical (the ceilometer points 4° off) and because of this it is highly sensitive to the presence of flat, plate-like ice crystals which typically fall with their longest axes aligned horizontally due to aerodynamic forces. These oriented crystals act like tiny mirrors, producing strong specular reflections of the laser beam back to the telescope, and this is measured by the lidar as a very large backscatter (but unlike liquid cloud, there is very little attenuation). This phenomenon turns out to be rather common in ice falling out of supercooled liquid cloud layers (Westbrook *et al.* 2009), particularly at temperatures between -10 and -20°C where plate-like crystals grow big and flat in the vapour-rich environment. The highly reflective virga of oriented crystals are visible between 5400 and 4700m. Note that the base of the fallstreak is around 700m lower than the natural virga either side of it, indicating that this dry air has been moistened sufficiently by the large flux of evaporating ice crystals that the crystals are able to survive the extra 700m of fall.

A photograph from the Chilbolton cloud camera (Figure 3) at 0951UTC clearly shows the Rayleigh-Bénard like cells of liquid cloud as the air is cooled radiatively from the top, destabiling the air and producing gentle convective overturning. The ice virga however are almost invisible to the eye, accounting for the ceilometer's insensitivity to them.

Figure 3 shows a second photograph taken at 0956UTC: a circular hole in the cloud layer is clearly visible directly overhead. Note that a visible ice fallstreak can be seen beneath the hole – this indicates that many new ice particles have been nucleated and grown, substantially increasing the ice content and optical depth relative to the surrounding 'ambient' ice virga. The lidar ceilometer, which until this point had not detected any ice, shows a very strong reflection from this fallstreak beneath the hole (5000-4500m), with backscatter values comparable to those found in the liquid cloud. This backscatter is an order of magnitude larger than is typically observed in ice-phase cloud (Westbrook *et al.* 2009). Note the gap in the liquid layer as the hole drifted overhead: this is approximately 3000m wide. Similarly the backscatter from the Doppler lidar is now approximately ten times larger than before; again a clear gap can be seen in the liquid cloud layer. Note that because of the 4° offset in pointing angle, the Doppler lidar is sampling a slightly different cross section of the fallstreak and hole, hence the hole appears narrower in the Doppler lidar time series (4° corresponds to a 350m horizontal offset at 5000m altitude).

At the same time, the 10cm CAMRa radar (Kilburn *et al.* 2000) was dwelling at 45° elevation pointing to the West of Chilbolton (into the wind). The time series from this instrument is shown in Figure 2. At this much longer wavelength, the tiny cloud droplets are essentially invisible (radar reflectivity is proportional the particle mass squared), and the returns are dominated by ice crystals. Again we observe that prior to the hole-punch there was very little ice being produced, with peak reflectivities of ≈ -5dBZ (corresponding to an ice water content of around 0.01gm$^{-3}$, Hogan *et al.* 2006); note much of the virga is close to the background signal level produced by noise and ground clutter. The ice falling out the hole-punch is visible as a streak of enhanced reflectivity, with values approximately 10dB (one order of magnitude) larger in the the ice than in the surrounding virga. The CAMRa radar also sends out pulses at different polarisations, and the parameter $Z_{DR}$ is the ratio of the signal for waves polarised horizontally *vs.* those polarised at 45° from the horizontal plane (in dB logarithmic units). This ratio contains information on the orientation, aspect ratio and density of the ice crystals. In the bulk of the ice virga the radar reflectivity was too weak to accurately estimate $Z_{DR}$; however in the strongest part of the fallstreak (>0dBZ) $Z_{DR}$ was 1.2±0.3dB. This mean value confirms that horizontally oriented ice crystals were indeed present, since the horizontally polarised reflectivity is higher than that at 45° (randomly oriented crystals would have $Z_{DR}$=0dB). Laboratory experiments by Takahashi et al. (1991) show that ice crystals grown in supercooled conditions at -20°C favour a thick hexagonal plate form, with an aspect ratio in the range 2.1-2.6 to 1. Assuming that all of the crystals were horizontally oriented and composed of solid ice, a theoretical calculation can be made to estimate what $Z_{DR}$ ought to be (for details see appendix in Westbrook et al. 2009). The calculated values are 1.5-1.8dB, which suggests that the majority of crystals in the observed fallstreak (1.2dB) were well oriented, and have similar properties to those grown in the lab.

**Origin of the fallstreak ice crystals**
As discussed by Pedgley (2008) the hypotheses that the crystals are produced by aerosol particles emitted by the aircraft exhaust has largely been discounted, due to the observations that a wide variety of aircraft burning different types of fuel are equally able to produce hole punches, and that they have been observed at temperatures as warm as -7°C (Rangno and Hobbs 1983,1984) where very few substances are able to act as active ice nuclei (Pruppacher and Klett 1997)

The generally accepted explanation is that hole-punches are produced by 'aerodynamic contrails', ie. trails of ice crystals produced in the rapidly expanding air at the tips of the wings (and flaps, gears, propellers if present). As the air expands it cools to ≈-40°C for a brief instant before mixing with the ambient air; at this very cold temperature liquid water droplets are able to freeze very rapidly without the aid of an aerosol particle (homogeneous freezing). However there is a subtlety: it is unclear whether (a) it is the droplets in the cloud which are frozen as they pass through this expanding air, or (b) whether new droplets are also formed in the expansion, which then go on to freeze. Both mechanisms were highlighted by Pedgley (2008): here our lidar and radar observations give some insight into which mechanism is likely dominant.

It seems that the hole-punch fallstreak in our observations contained a substantially enhanced population of ice crystals relative to the natural virga, and this is borne out by the fact that it is so visible in the cloud camera photograph. The enhancement in the lidar backscatter and radar reflectivity is a factor of 10. The implication therefore is that the crystal concentration is 10× larger. We first consider mechanism (a). A typical altocumulus cloud has around 200$cm^{-3}$ droplets (Hobbs and Rangno 1985) – these would be frozen in a thin ribbon behind the wing tips, each perhaps only 10cm wide. However, by the time we observed the fallstreak it was approximately 500m across – thus the concentration of crystals at that point in time must have been diluted by $500^2/(2\times0.1^2)$, approximately 10 million-fold. This means that only around 0.02 crystals per litre of air would be present in the fallstreak – compare that to typical concentrations of crystals in natural altocumulus virga of 1-10 per litre (Hobbs and Rangno 1985). So the enhancement of the lidar backscatter and radar reflectivity relative to the surrounding virga would be insignificant; nor would the fallstreak be so visually striking in the cloud camera photograph. We therefore discount mechanism (a).

Our observations point to mechanism (b) as the dominant source of ice crystals. As the air expands and cools rapidly behind the propellor blades and wing tips, new water droplets are formed in the process. Figure 4 shows an example of a trail of liquid droplets being formed at the tips of an aircraft's propellers by this process whilst flying at warm temperatures. Because the saturation vapour pressure for liquid water is ten times lower at -40°C compared to the ambient air at -20°C, an enormous (but very localised) relative humidity of almost 800% is possible immediately following the expansion of the water-saturated air in the liquid cloud layer. This allows liquid droplets to form *homogeneously* (directly from vapour, without an aerosol particle) – something that never occurs naturally in the atmosphere (Mason 1971), but which is possible here because of the extremely rapid expansion (Maybank and Mason 1951). The production rate of new droplets via this mechanism may be calculated (following Foster and Hallett 2003), and we find that it is a prolific source ($\approx 10^{14} m^{-3} s^{-1}$) of new, miniscule droplets, which then freeze immediately, since at -40°C the rate of homogeneous freezing is even faster than the droplets are being produced (Pruppacher and Klett 1997). Even assuming that the air maintains this high humidity for a mere 0.01 seconds after the air is cooled before mixing sets in, $10^{12} m^{-3}$ new ice crystals would be formed in the condensation ribbon; diluting this to a 500m width still leaves 100 ice crystals per litre in the fallstreak, an order of magnitude or more higher than the natural virga concentration, and easily capable of explaining our observed lidar and radar measurements. This explanation has also been highlighted as most likely by Woodley et al. (1991,2003), Foster and Hallett (1993) and Vonnegut (1986). Truly then, hole-punch clouds are an extreme example of nucleation and glaciation, and in fact the mechanism described above is essentially identical to that which occurs when clouds are seeded using dry ice, the surface of which is so cold (-78°C) that the air in the immediate vicinity is rapidly cooled to below -40°C for a short period, again producing numerous tiny crystals via homogeneous condensation and freezing (Mason 1981).

**Mirror reflections**
This mechanism for producing the ice crystals also likely explains the strong specular reflection which was observed. Large cloud droplets frozen at cold temperatures develop multiple crystallographic directions and grow into complex polycrystals (e.g. Bacon *et al.* 2003) which would not produce the observed mirror reflections. However polycrystal development is a strong function of droplet size (Pitter and Pruppacher 1973): if tiny droplets are formed homogeneously and freeze before having a chance to grow to 10μm or more, they will freeze into single hexagonal crystals, and then grow into simple planar forms at the ambient cloud temperature of -20°C. Experimental support for this is idea is provided by

Mason (1953) who found that simple hexagonal-type crystals are produced in supercooled clouds seeded with dry ice at temperatures as cold as -39°C: we expect ice crystals from hole-punch clouds to follow the same behaviour. Similarly, Rangno and Hobbs (1983,1984) sampled the ice particle fallstreaks produced when their aircraft penetrated altocumulus layers and found large concentrations of simple hexagonal columns at -7°C and hexagonal plates at -11°C. This idea also offers an explanation for Pedgley's difficulty in explaining observations of a mock sun produced by a hole-punch fallstreak at a temperature of -24°C. This optical phenomenon is also the result of mirror reflections (of sunlight) from horizontally oriented plate-like crystals. Again, we expect that the crystals grew in a plate-like manner typical of the ambient temperature (Bacon *et al.* 2003, Bailey and Hallett 2004), but whereas cloud droplets might grow into complex polycrystals, the tiny initial droplets produced here would grow into simple thick plates, which tend to orient horizontally as they fall (Podzimek 1968) and reflect the sunlight as observed.

**Vertical air motions**
The vertical velocities measured by the Doppler lidar in Figure 2 are very intriguing. Positive values indicate particles ascending away from the lidar, negative values indicate particles falling toward the lidar. In the liquid layer we see small up and down motions corresponding to the gentle convective overturning at cloud top. Directly beneath the hole the crystals are seen to be falling at a few tens of centimetres per second: such values are typical of small planar crystals settling with their major axis horizontal (Pruppacher and Klett 1997). However either side of the main fallstreak there appear to be vertical columns of rising air (ice crystals ascending at $0.2ms^{-1}$) throughout the depth of the virga, and more vigorous convective overturning in the cloud layer at the edge of the hole (cloud droplets rising at up to $0.5ms^{-1}$). It almost appears as though the fallstreak has caused a 'ripple' in the surrounding air. One possibility is that the enormous flux of ice crystals into the dry, stable air below has caused substantial evaporative cooling, and therefore turbulence. We speculate that this disturbance of the stable air layer produces oscillations which may then in turn disrupt the air in the cloud layer above, promoting condensation and mixing, and perhaps helps to reform the liquid cloud layer from the edges as the ice falls out. Alternatively, the large moisture flux may have been sufficient to destabilise the virga itself, leading to convective motion through the depth of the fallstreak and its immediate surroundings.

Do readers have any other observations or ideas on hole-punch clouds which might build on or disprove some of the ideas presented above? With the advent of cheap digital cameras with video capability it should now be possible to capture the evolution of a hole-punch over 10 minutes or so (if there's no low cloud in the way) – so if you're lucky enough to spot a hole, get filming!

**Acknowledgements**:
We would like to thank the staff at the STFC Chilbolton Observatory for operation and maintenance of the lidars and CAMRa radar. We would also like to thank Robin Hogan, Anthony Illingworth, Jon Eastment and our two reviewers for their helpful comments on the manuscript, and William Woodley for allowing us to use his photograph in Figure 4. This work was funded by the Natural Environment Research Council grants NER/Z/S/2003/00643 and NE/EO11241/1.

**References**:
**Bacon NJ, Baker MB, Swanson BD.** 2003. Initial stages in the morphological evolution of vapour-grown ice crystals: a laboratory investigation. *Q. J. R. Meteor. Soc.* **129**: 1903-1927
**Bailey M and Hallett J.** 2004. Growth rates and habits of ice crystals between -20 and -70°C. *J. Atmos. Sci.* **61**: 514-544
**Foster TC and Hallett J.** 1993. Ice crystals produced by expansion: experiments and application to aircraft-produced ice. *J. Atmos. Sci.* **32**: 716-728.
**Hobbs PV and Rangno AL**. 1985. Ice particle concentrations in clouds *J. Atmos. Sci.* **42**: 2523-2549
**Kilburn CAD, Chapman D, Illingworth AJ and Hogan RJ.** 2000. Weather observations from the Chilbolton Advanced Meteorological Radar. *Weather,* **55** 352-355.
**Mason** BJ. 1953. The growth of ice crystals in a supercooled water clouds. *Q. J. R. Meteor. Soc.* **79** 105-111

**Mason** BJ. 1981. The mechanisms of seeding with dry ice. *J. Wea. Mod.* **13** 11
**Maybank J and Mason BJ** 1951. Spontaneous condensation of water vapour in expansion chamber experiments *Proc. Phys. Soc. B* **64** 773-779
**Pedgley DE.** 2008. Some thoughts on fallstreak holes. *Weather* **63** 356-360.
**Podzimek J.** 1968. Aerodynamic conditions of ice crystal aggregation. *Int. Conf. Cloud Phys. (Toronto) - proceedings* 295-298.
**Pitter RL and Pruppacher HP.** 1973. A wind tunnel investigation of freezing of small water drops falling at terminal velocity in air *Q. J. R. Meteor. Soc.* **99** 540-550
**Pruppacher H and Klett JD.** 1997. *Microphysics of Clouds and Precipitation* Kluwer.
**Rangno AL and Hobbs PV.** 1983. Production of ice particles in clouds due to aircraft penetrations. *J. Appl. Met.* **22**: 214-232
**Rangno AL and Hobbs PV.** 1984. Further observations of the production of ice particles in clouds by aircraft *J. Clim. & Appl. Met.* **23**: 985-987
**Takahashi T, Endoh T, Wakahama G.** 1991. Vapor diffusional growth of free-falling snow crystals between -3 and -23°C *J. Met. Soc. Japan* **69**: 15
**Vonnegut B.** 1986. Nucleation of ice crystals in supercooled clouds caused by passage of an airplane *J. Clim. & Appl. Met.* **25**: 98
**Westbrook CD, Illingworth AJ, O'Connor EJ and Hogan RJ.** 2009. Doppler lidar measurements of oriented planar ice crystals falling from supercooled and glaciated layer clouds. Accepted *Q. J. R. Meteor. Soc.*
**Woodley WL, Henderson TJ, Vonnegut B, Gordon G, Rosenfeld D, Holle SM**. 1991. Aircraft produced ice particles (APIPs) in supercooled clouds and the probably mechanism for their production. *J. Appl. Met.* **30**: 1469-1489
**Woodley WL, Gordon G, Henderson TJ, Vonnegut B, Rosenfeld D, Detwiler A**. 2003. Aircraft produced ice particles (APIPs): additional results and further insights. *J. Appl. Met.* **42**: 640-651

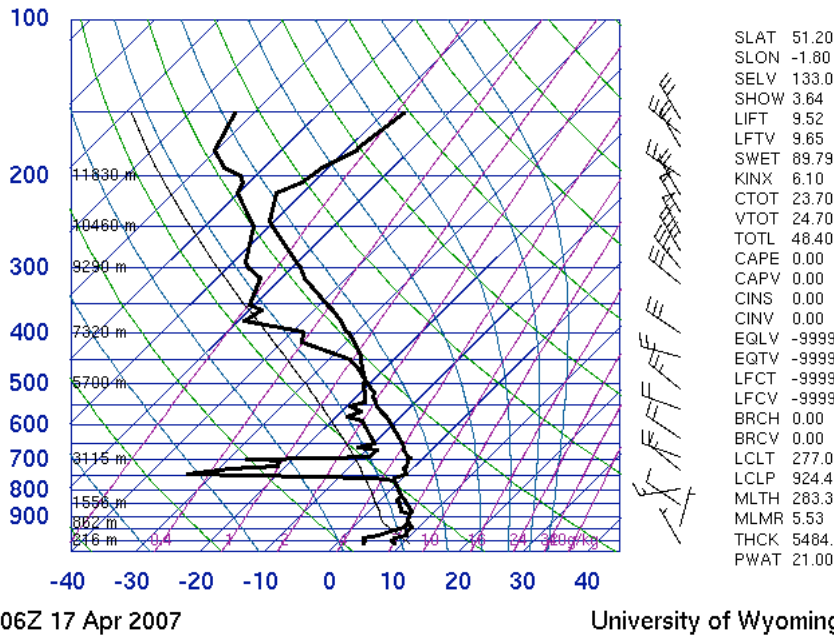

**Figure 1:** Skew-T diagram showing the radiosonde ascent from Larkhill (25km West of Chilbolton) at 0600UTC, courtesy of the Department of Atmospheric Sciences, University of Wyoming. Note the liquid water saturated layers at 1000, 2000 and 5800m.

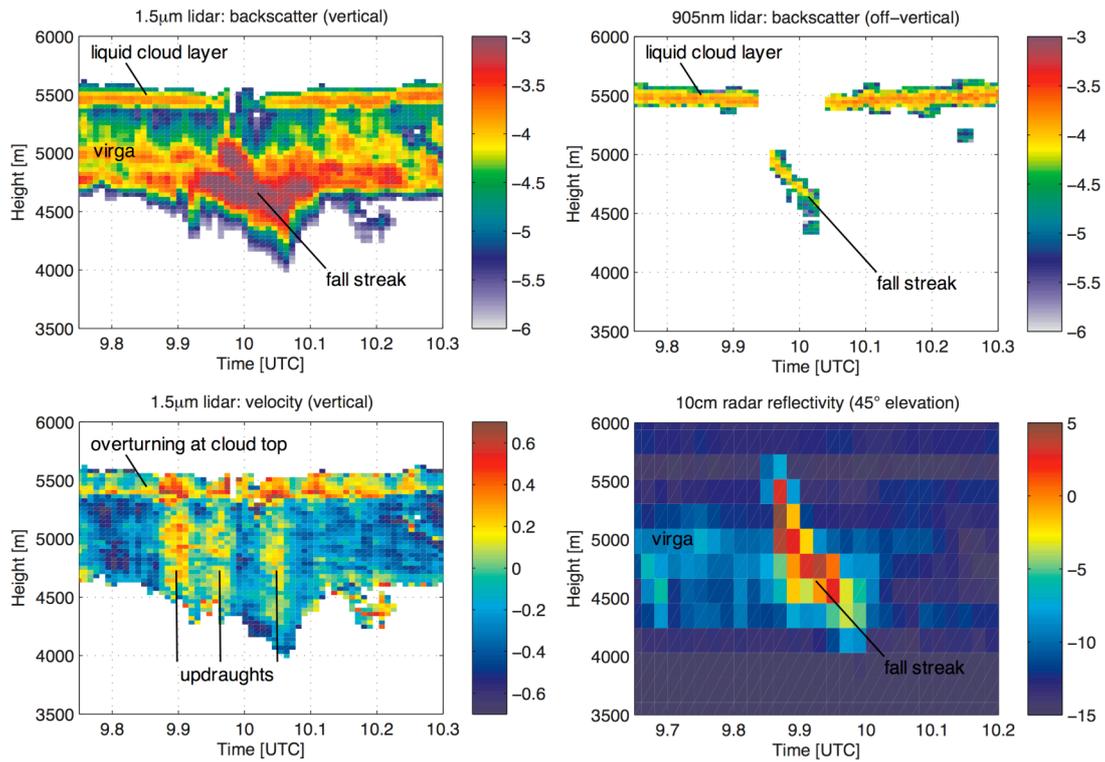

**Figure 2:** Lidar and radar time series. Top left shows backscatter from vertically pointing Doppler lidar; top right shows backscatter from lidar ceilometer pointing 4° away from vertical. Bottom right shows radar reflectivity measured while dwelling at 45° elevation due West. Bottom left shows vertical velocity measurements from the Doppler lidar. Lidar backscatter colour scales are logarithmic = $\log_{10}[m^{-1}sr^{-1}]$, radar reflectivity colour scale is in $dBZ = 10\log_{10}[mm^6 m^{-3}]$. Doppler velocity scale is in $ms^{-1}$, positive values indicate particles moving away from the lidar (ie. upwards).

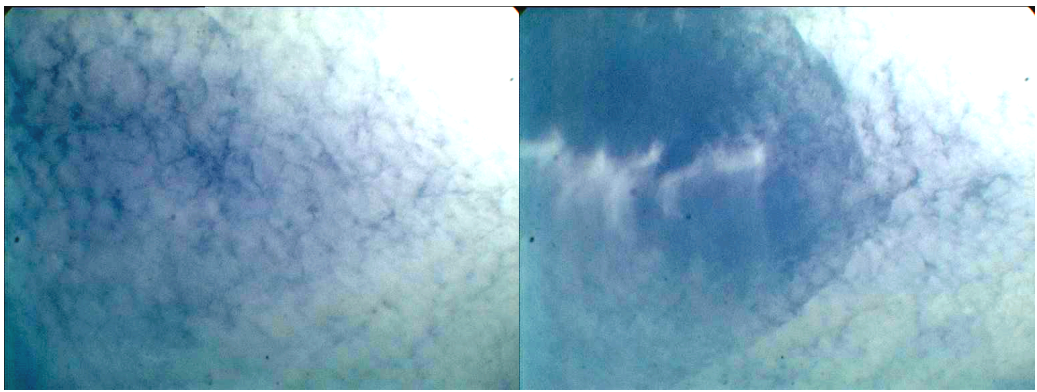

**Figure 3:** Two consecutive photographs of the sky from the Chilbolton cloud camera. Left panel was taken at 0951UTC; right panel was taken 5 minutes later and shows a large circular hole in the altocumulus layer with a highly reflective ice fallstreak beneath.

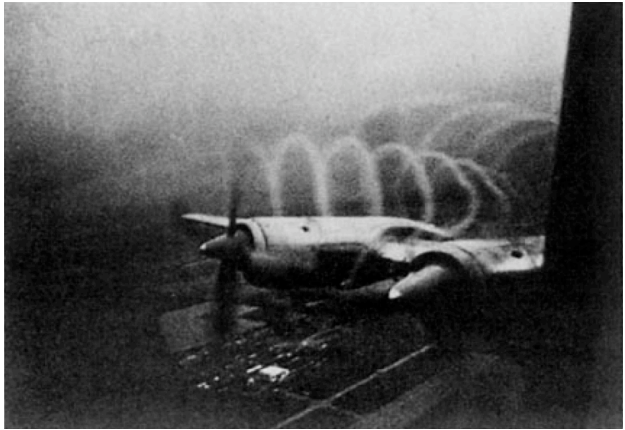

**Figure 4:** Thin helical ribbon of condensation formed in the rapidly cooled air behind the propellor tips of an aircraft. Photograph courtesy William L. Woodley.